\documentclass[12pt]{article}
\usepackage{graphicx}

\newcommand{\beq}{\begin{equation}}
\newcommand{\eeq}{\end{equation}}
\newcommand{\beqa}{\begin{eqnarray}}
\newcommand{\eeqa}{\end{eqnarray}}
\newcommand{\non}{\nonumber}
\newcommand{\lab}{\label}

\newcommand{\ket}{\rangle}

\begin{document}

\title{A quantum algorithm for examining oracles}

\author{Hiroo Azuma \\
{\small Canon Inc., 5-1, Morinosato-Wakamiya,}\\
{\small Atsugi-shi, Kanagawa, 243-0193, Japan}\\
{\small E-mail: azuma.hiroo@canon.co.jp}}

\date{October 7, 2004}

\maketitle

\begin{abstract}
In this paper, we consider a quantum algorithm
for solving the following problem:
``Suppose $f$ is a function given as a black box
(that is also called an oracle)
and $f$ is invariant under some AND-mask.
Examine a property of $f$ by querying the oracle.''
We compare the efficiency of our quantum algorithm with that of classical algorithms
by evaluating the expected number of queries for each algorithm.
We show that our quantum algorithm is more efficient than any classical algorithm in some cases.
However, our quantum algorithm does not exhibit an exponential speedup in the size of an input,
compared with the best classical algorithm.
Our algorithm extracts a global property of $f$
(that is, invariance of $f$)
while it neglects local properties of $f$
(that is, outputs of $f$).
We can regard our algorithm as an application of Simon's algorithm.
\end{abstract}

\section{Introduction}
\lab{introduction}
Since R.P.~Feynman claimed in 1980's that we need a computer that runs on the principle of quantum mechanics
to simulate a quantum system efficiently,
many researchers have been studying quantum computation \cite{Feynman}.
After his early work,
D.~Deutsch formalized the current model of the quantum computation \cite{Deutsch}.
Because it is believed that a quantum computer solves certain problems faster than any classical computer,
the quantum computation draws many researchers' attention.
Well-known examples of quantum algorithms are
Deutsch and Jozsa's algorithm, Simon's algorithm,
Shor's algorithm, and Grover's algorithm \cite{Deutsch-Jozsa,Simon,Shor,Grover}.
(Shor's algorithm shows that a quantum computer can factor integers and find discrete logarithms
in polynomial time.
It is widely believed that no classical algorithm solves these problems
in polynomial time.
Comprehensive reviews of the quantum computation are given in Ref.~\cite{QuantumComputationReview}.)
These algorithms make good use of the properties of quantum mechanics, namely, 
the principle of superposition and its interference,
entanglement, and the principle of uncertainty.

Simon's algorithm solves the following problem.
Suppose $f:\{0,1\}^{n}\rightarrow \{0,1\}^{n}$ is a function
given as a black box, which is called an oracle by computer scientists.
We are promised that one of two cases occurs:
(1) all $f(x)$ (for $x\in\{0,1\}^{n}$) are different, that is,
$f$ is one-to-one;
(2) there exists an unknown $n$-bit string $s(\neq\mbox{\boldmath$0$})$
such that $\forall x,y\in\{0,1\}^{n}$ ($f(x)=f(y)$ if and only if $x=y$ or $x=y\oplus s$), that is,
$f$ is two-to-one.
Determine which of the two cases holds, and in the second case,
find $s$.
($\mbox{\boldmath$0$}$ is an $n$-bit string whose every bit has a zero
in its entry, $\mbox{\boldmath$0$}=0...0$.
$\oplus$ denotes the bitwise XOR.)

This problem is called Simon's problem.
We want to solve it as efficiently as possible.
(We want to let the number of queries to the oracle become fewer.)
To solve the problem by a classical method,
we need at least of the order of $2^{n}$ queries on average.
(To be exact, any classical bounded-error probabilistic computer requires exponential time
in the expected sense
to solve Simon's problem.)
By contrast, Simon's quantum algorithm solves the problem with $poly(n)$ queries
on average, where $poly(n)$ denotes a polynomial in $n$.

Simon's algorithm finds a period of the function $f$.
It neglects local information of $f$, that is,
which value $f(x)$ takes for each $x\in\{0,1\}^{n}$,
and extracts only global information of $f$, that is,
the period $s$.
Shor's algorithm has this feature as well.

Developing a new quantum algorithm is an important topic in the field of the quantum computation.
In this paper, we consider the following problem that is similar to Simon's.
Let us assume that the function $f:\{0,1\}^{n}\rightarrow \{0,1\}^{n}$
is given as an oracle and $f$ has invariance $f(x\wedge s)=f(x)$.
We also assume that we do not know the $n$-bit string $s$
except that the Hamming weight of $s$ is given by $\mbox{wt}(s)=m$.
The problem is to find $s$.
($\wedge$ denotes the bitwise AND.
We assume $1\leq m\leq n-1$.
As shown later, $f$ is a $2^{n-m}$-to-one function.)

We want to solve this problem as efficiently as possible, too.
Evaluating the lower bound of the expected number of queries
required by a classical computer to solve the problem,
we show that it is given by
$1+\lceil \log_{2}n \rceil$ for $\mbox{wt}(s)=1$
and it is given by
$-(1/n)+(3/2)+(n/2)$
for $\mbox{wt}(s)=n-1$,
respectively.
In the case of $2\leq \mbox{wt}(s)\leq n-2$,
we cannot derive the classical lower bound.

By contrast, the expected number of queries for our quantum algorithm
approximates $2+\log_{2}m$, where $m=\mbox{wt}(s)$.
Furthermore we note that the expected number of queries for $\mbox{wt}(s)=1$
is equal to exactly two.
Thus we conclude that our quantum algorithm is more efficient
than any classical algorithm
for $\mbox{wt}(s)=1$ and $\mbox{wt}(s)=n-1$.

A network of quantum gates for our algorithm is the same as that for Simon's algorithm.
Hence, we can regard our algorithm as an application of Simon's.
However, ours has the following new feature.
To obtain the binary string $s$, we apply a certain classical procedure to values
observed in trials of our algorithm.
Thus we have to use a recurrence formula to derive the expected number of queries.
This point cannot be found in the other quantum algorithms.

A good quantum algorithm (like Simon's) takes polynomial time
in $n$ for solving a certain problem,
where $n$ is the size of an input,
while any classical algorithm takes exponential time in $n$.
But, our algorithm does not exhibit such an exponential quantum speedup in the size of an input,
compared with the most efficient classical algorithm.
However, our quantum algorithm extracts only the global property of the oracle
with neglecting the local properties of the oracle.
We think this fact important for understanding quantum computation.

This paper is organized as follows.
In Sec.~\ref{problem-quantum-algorithm},
we define a problem that we consider through this paper and describe a quantum algorithm to solve it.
We explain how this algorithm works.
In Sec.~\ref{queries-quantum-algorithm},
we estimate the expected number of queries to the oracle in the quantum algorithm defined
in Sec.~\ref{problem-quantum-algorithm}.
In Sec.~\ref{queries-classical-algorithm-lower-limit},
we discuss the lower bound of the number of queries required by any classical algorithm
for solving the problem defined in Sec.~\ref{problem-quantum-algorithm}.
We evaluate the lower bounds for $\mbox{wt}(s)=1$ and $\mbox{wt}(s)=n-1$.
In Sec.~\ref{queries-classical-algorithm-between-2-(n-2)},
we investigate two typical classical algorithms for $2\leq\mbox{wt}(s)\leq n-2$.
We estimate the expected number of queries for each of them.
In Sec.~\ref{OR-quantum-algorithm},
we consider a quantum algorithm for examining a function that is invariant under some OR-mask.
In Sec.~\ref{discussions},
we give a brief discussion.

\section{A problem and a quantum algorithm}
\lab{problem-quantum-algorithm}
In this section, we define a problem that we discuss through this paper,
and we give a quantum algorithm to solve it.
Then, we explain how our algorithm works.

First of all, we give some notations.
We prepare two arbitrary $n$-bit strings
$a,b\in\{0,1\}^{n}$.
(We have $a=a_{1}a_{2}...a_{n}$, where $a_{i}\in\{0,1\}$ for $i=1,...,n$.
We apply the same to $b$.)
We define the bitwise AND of $a$ and $b$ as $c=a\wedge b\in\{0,1\}^{n}$,
where $c_{i}=a_{i}b_{i}$ for $i=1,...,n$.
We define the bitwise OR of $a$ and $b$ as $c=a\vee b\in\{0,1\}^{n}$,
where $c_{i}=\mbox{max}(a_{i},b_{i})$ for $i=1,...,n$.
We define the bitwise XOR of $a$ and $b$ as $c=a\oplus b\in\{0,1\}^{n}$,
where $c_{i}=a_{i}+b_{i} \pmod{2}$ for $i=1,...,n$.
Moreover, we write the inner product of $a$ and $b$ as $a\cdot b=\sum_{i=1}^{n}a_{i}b_{i} \pmod{2}$.
We describe the number of nonzero bits in $a$ as $\mbox{wt}(a)$,
and we call it the Hamming weight of $a$.
(Clearly it satisfies $0\leq\mbox{wt}(a)\leq n$.)
$\overline{a}$ is a binary string obtained by reversing each bit of $a$ as $0\leftrightarrow 1$.
We write an $n$-bit string whose every bit is equal to zero as $\mbox{\boldmath $0$}(=0...0)$,
and an $n$-bit string whose every bit is equal to one as $\mbox{\boldmath $1$}(=1...1)$.
(We obtain relations, $\mbox{wt}(\mbox{\boldmath $0$})=0$,
$\mbox{wt}(\mbox{\boldmath $1$})=n$,
$\overline{\mbox{\boldmath $0$}}=\mbox{\boldmath $1$}$,
and $\overline{\mbox{\boldmath $1$}}=\mbox{\boldmath $0$}$,
immediately.)

Let us consider a state on a system that consists of qubits.
(The qubit is a two-state system $\{|0\ket,|1\ket\}$.)
Quantum computation is a sequence of
unitary transformations and measurements applied to this multi-qubit system.
We define the following two unitary transformations.
The first one is the Hadamard transformation $H$,
which works on a qubit as follows:
\beq
H:
\left\{
\begin{array}{lll}
|0\ket & \rightarrow & (1/\sqrt{2})(|0\ket+|1\ket) \\
|1\ket & \rightarrow & (1/\sqrt{2})(|0\ket-|1\ket)
\end{array}
\right..
\lab{definition-H}
\eeq
$H^{\otimes n}$ transforms an $n$-qubit state
$|x\ket=|x_{1}\ket...|x_{n}\ket$ $\forall x\in\{0,1\}^{n}$
as follows:
\beq
H^{\otimes n}|x\ket=\frac{1}{\sqrt{2^{n}}}\sum_{y\in\{0,1\}^{n}}(-1)^{x\cdot y}|y\ket.
\lab{action-H-n-qubits}
\eeq
The second one is $U_{f}$
that realizes an oracle of a function
$f:\{0,1\}^{n}\rightarrow\{0,1\}^{n}$
as follows:
\beq
U_{f}|x\ket|y\ket=|x\ket|y+f(x)\pmod{2^{n}}\ket
\quad\quad
\forall x,y\in\{0,1\}^{n}.
\eeq

We define the problem as follows:

\bigskip
\noindent
[Problem]

\noindent
Suppose that we are given a function
$f:\{0,1\}^{n}\rightarrow \{0,1\}^{n}$.
We are promised that there exists an $n$-bit string $s$ such that
$\forall x,y\in\{0,1\}^{n}$
($f(x)=f(y)$ if and only if $x\wedge s=y\wedge s$).
We do not know $s$ except that we are given $\mbox{wt}(s)=m$.
Find $s$.

\bigskip

In this problem, we find immediately that
$s=\mbox{\boldmath$0$}$ if $\mbox{wt}(s)=0$
and
$s=\mbox{\boldmath$1$}$ if $\mbox{wt}(s)=n$.
Thus, we assume $s\neq \mbox{\boldmath$0$},\mbox{\boldmath$1$}$.
As shown later, $f$ is a $2^{n-m}$-to-one function.

We consider the following quantum algorithm.

\bigskip
\noindent
[Algorithm]

\noindent
\begin{enumerate}
\item We prepare two registers that consist of $n$ qubits respectively,
and put each qubit in $|0\ket$ as an initial state.
We obtain a state $|0\ket^{n}|0\ket^{n}$.
\item We apply the Hadamard transformation $H$ to each qubit of the first register.
From Eq.~(\ref{action-H-n-qubits}), we obtain
\beq
\frac{1}{\sqrt{2^{n}}}\sum_{x\in\{0,1\}^{n}}|x\ket|0\ket.
\eeq
(Here, we rewrite $|0\ket^{n}$ on the second register as $|0\ket$ for simplicity.)
\item Applying the given oracle $U_{f}$ to the registers,
we store $f(x)$ in the second register according to the value $x$ in the first register.
Thus, we obtain
\beq
\frac{1}{\sqrt{2^{n}}}\sum_{x\in\{0,1\}^{n}}|x\ket|f(x)\ket.
\lab{quantum-algorithm-3rd-step-register}
\eeq
\item We apply $H$ to each qubit of the first register again.
We obtain
\beq
\frac{1}{\sqrt{2^{n}}}\sum_{x\in\{0,1\}^{n}}(H^{\otimes n}|x\ket)|f(x)\ket.
\eeq
\item We observe the first register in a logical ket basis $\{|x\ket:x\in\{0,1\}^{n}\}$.
Let us suppose that we obtain $|k\ket$.
If $\mbox{wt}(k)=m$, we let $s=k$.
If $\mbox{wt}(k)<m$, we carry out operation from the first step to the fourth step again
and observe the first register.
(We call this process a trial.)
Rewriting $k$ obtained by the first trial as $k_{old}$
and writing a string obtained by the second trial as $k_{new}$,
we calculate
\beq
k=k_{old}\vee k_{new}.
\lab{OR-operation}
\eeq
If $\mbox{wt}(k)=m$, we let $s=k$.
If $\mbox{wt}(k)<m$, we rewrite $k$ as $k_{old}$,
obtain $k_{new}$ by another trial,
and calculate $k=k_{old}\vee k_{new}$.
We repeat this procedure until we have $\mbox{wt}(k)=m$.
When $\mbox{wt}(k)=m$, we obtain $s=k$.
\end{enumerate}

We explain the reason why we can obtain $s$ by the above algorithm.
First, we pay attention to the following fact.
$\forall x,y\in\{0,1\}^{n}$, $f(x)=f(y)$ if and only if $x\wedge s=y\wedge s$,
and $\mbox{wt}(s)=m$ is given.
Thus, $f$ is a $2^{n-m}$-to-one function.
This is because the number of bits that hold zeros in the string $s$ is
equal to $(n-m)$ and the function $f$ does not depend on these $(n-m)$ bits.
Hence, we can classify $2^{n}$ inputs of $f$ (that is, $x\in\{0,1\}^{n}$)
into $2^{m}$ classes according to values that they take as outputs of $f$ (that is, $f(x)$).
The number of inputs in each class is equal to $2^{n-m}$.

From this consideration, we can rewrite Eq.~(\ref{quantum-algorithm-3rd-step-register})
that is obtained after the third step of the algorithm as follows:
\beq
\frac{1}{\sqrt{2^{n}}}
\sum_{l:l=x\wedge s,x\in\{0,1\}^{n}}
(\sum_{a:a=y\wedge \overline{s},y\in\{0,1\}^{n}}
|a\oplus l\ket)|f(l)\ket.
\lab{Q-algorithm-3rd-step}
\eeq
In Eq.~(\ref{Q-algorithm-3rd-step}), the binary string $l$ has zeros in entries where $s$ has zeros,
and $l$ has either zeros or ones at random in entries where $s$ has ones.
Thus, there are $2^{m}$ possible strings for $l$.
Meanwhile the binary string $a$ has zeros in entries where $\overline{s}$ has zeros,
and $a$ has either zeros or ones at random in entries where $\overline{s}$ has ones.
(Put another way,
$a$ has zeros in entries where $s$ has ones,
and $a$ has either zeros or ones at random in entries where $s$ has zeros.)
Thus, there are $2^{n-m}$ possible strings for $a$.

Next, we apply $H$ to each qubit of the first register in Eq.~(\ref{Q-algorithm-3rd-step})
for the fourth step.
Here, we consider only the $n$ qubits of the first register,
$(1/\sqrt{2^{n-m}})\sum_{a}|a\oplus l\ket$.
We permute these $n$ qubits, so that zeros of the string $s$ move to the left side and ones of $s$
move to the right side.
Because $H$ works upon each qubit independently,
this permutation does not change the essence of this discussion.
By this permutation, we rewrite $(1/\sqrt{2^{n-m}})\sum_{a}|a\oplus l\ket$ as
\beq
\frac{1}{\sqrt{2^{n-m}}}\sum_{a'\in\{0,1\}^{n-m}}|a'\ket|l'\ket,
\lab{Q-algorithm-3rd-step-1st-register}
\eeq
where we can obtain $l'$ by permuting the $n$-bit string $l$ and removing
$(n-m)$ zeros from the left side of its entries.
Thus, $|a'\ket$ is an $(n-m)$-qubit state
and $|l'\ket$ is an $m$-qubit state.

We apply $H^{\otimes n}$ to the state of Eq.~(\ref{Q-algorithm-3rd-step-1st-register}).
From Eq.~(\ref{action-H-n-qubits}), we find that it transforms the state of the first
$(n-m)$ qubits
$(1/\sqrt{2^{n-m}})\sum_{a'}|a'\ket$
to $|0\ket^{n-m}$.
Hence, the state of Eq.~(\ref{Q-algorithm-3rd-step-1st-register}) is transformed to
\beq
\frac{1}{\sqrt{2^{m}}}\sum_{k'\in\{0,1\}^{m}}(-1)^{l'\cdot k'}|0\ket^{n-m}|k'\ket.
\lab{Q-algorithm-4th-step-permuted}
\eeq
Permuting the qubits in Eq.~(\ref{Q-algorithm-4th-step-permuted}) to the original order,
we obtain the following state on both the registers:
\beq
\frac{1}{2^{m}}
\sum_{l:l=x\wedge s,x\in\{0,1\}^{n}}
[\sum_{k:k=y\wedge s,y\in\{0,1\}^{n}}
(-1)^{l\cdot k}|k\ket]|f(l)\ket.
\lab{Q-algorithm-4th-step}
\eeq

Then, for the fifth step, we observe the first register in the basis $\{|x\ket:x\in\{0,1\}^{n}\}$.
We obtain a binary string $k$,
where $k=y\wedge s$ and $y\in\{0,1\}^{n}$.
There are $2^{m}$ possible strings for $k$,
and $k$ takes one of them at random.
$k$ has zeros in entries where $s$ has zeros,
and $k$ has either zeros or ones at random in entries where $s$ has ones.
Thus, if we repeat the trial with observing a string $k$ and perform the bitwise OR
to observed strings $k$ as Eq.~(\ref{OR-operation}) again and again,
we will obtain $s$ eventually.
In Eq.~(\ref{Q-algorithm-4th-step}),
we find that bits in the first register depend on only $k$
and phases include information of $l$.

Here, let us see a concrete example of our algorithm.
We suppose $n=3$, $s=110$, and $\mbox{wt}(s)=2$.
Then, $f$ is two-to-one.
Thus, we may define $f:\{0,1\}^{3}\rightarrow\{0,1\}^{2}$.
We can classify inputs $x\in\{0,1\}^{3}$ to the following four classes:
\beqa
f(000)=f(001),& \quad & f(010)=f(011), \non \\
f(100)=f(101),& \quad & f(110)=f(111).
\lab{classification-inputs}
\eeqa

Preparing an initial state $|000\ket|00\ket$ and applying $H^{\otimes 3}$ to the first register,
we obtain
\beq
\frac{1}{\sqrt{8}}\sum_{x\in\{0,1\}^{3}}|x\ket|0\ket.
\lab{a-uniform-superposition-of-3-qubits}
\eeq
Applying the oracle $U_{f}$ to Eq.~(\ref{a-uniform-superposition-of-3-qubits}), we obtain
\beqa
&&(1/\sqrt{8})\sum_{x}|x\ket|f(x)\ket \non \\
&=&
(1/\sqrt{8})[(|000\ket+|001\ket)|f(000)\ket+(|010\ket+|011\ket)|f(010)\ket \non \\
&&\quad +(|100\ket+|101\ket)|f(100)\ket+(|110\ket+|111\ket)|f(110)\ket].
\eeqa
Applying $H^{\otimes 3}$ to the first register again,
we obtain
\beqa
&&(1/4)[(|000\ket+|010\ket+|100\ket+|110\ket)|f(000)\ket \non \\
&&\quad +(|000\ket-|010\ket+|100\ket-|110\ket)|f(010)\ket \non \\
&&\quad +(|000\ket+|010\ket-|100\ket-|110\ket)|f(100)\ket \non \\
&&\quad +(|000\ket-|010\ket-|100\ket+|110\ket)|f(110)\ket].
\eeqa

If we observe the first register, we obtain
$|000\ket$, $|010\ket$, $|100\ket$, or $|110\ket$ at random.
Let us assume that we obtain a string $k^{(1)}=010$ in the first trial.
We know $\mbox{wt}(s)=2$ beforehand.
Because $\mbox{wt}(k^{(1)})=1$,
we find $s\neq k^{(1)}$.
Then, let us suppose that we obtain a string $k^{(2)}=100$ in the second trial.
Calculating $k=k^{(1)}\vee k^{(2)}=110$ and noticing $\mbox{wt}(k)=2$,
we find $s=k=110$.

\section{The expected number of queries for the quantum algorithm}
\lab{queries-quantum-algorithm}
In this section, we evaluate the expected number of queries required by the quantum algorithm
shown in Sec.~\ref{problem-quantum-algorithm}.
Moreover, we discuss some features of our algorithm.

For a start, we investigate running time for our algorithm.
As shown later, the expected number of queries depends on the Hamming weight of the string $s$
(that is, $\mbox{wt}(s)=m$),
while it does not depend on the number of qubits $n$.
Thus, we describe it as $T_{\mbox{\scriptsize Q}}(m)$.
The subscript Q of $T_{\mbox{\scriptsize Q}}(m)$ stands for ``quantum''.

Let us evaluate $T_{\mbox{\scriptsize Q}}(1)$.
(We are given $\mbox{wt}(s)=m=1$.)
We can assume $s=10...0$ without losing generality.
From Eq.~(\ref{Q-algorithm-4th-step}), after the fourth step of our algorithm,
we obtain
\beq
\frac{1}{2}[(|0...0\ket+|10...0\ket)|f(0...0)\ket
+(|0...0\ket-|10...0\ket)|f(10...0)\ket].
\lab{final-state-m=1-case} 
\eeq

We observe the first register.
We carry out the trial again if we obtain $|0...0\ket$,
and we finish the task if we obtain $|s\ket=|10...0\ket$.
Thus, we can write $T_{\mbox{\scriptsize Q}}(1)$ as
\beq
T_{\mbox{\scriptsize Q}}(1)
=1\cdot\frac{1}{2}+2\cdot(\frac{1}{2})^{2}+3\cdot(\frac{1}{2})^{3}+...
=\sum_{h=1}^{\infty}h(\frac{1}{2})^{h}.
\eeq
Using the formula
\beqa
\sum_{h=1}^{\infty}hx^{h}
&=& x\frac{d}{dx}\sum_{h=1}^{\infty}x^{h}=x\frac{d}{dx}\frac{x}{1-x} \non \\
&=& \frac{x}{(1-x)^{2}} \quad\mbox{for $|x|<1$},
\eeqa
we obtain $T_{\mbox{\scriptsize Q}}(1)=2$.
Here, we notice that $T_{\mbox{\scriptsize Q}}(1)$ does not depend on $n$.

We can also derive $T_{\mbox{\scriptsize Q}}(1)$ by another way as follows.
If we observe the state given by Eq.~(\ref{final-state-m=1-case}),
we obtain either $|0...0\ket$ or $|s\ket=|10...0\ket$ with probability $1/2$ respectively.
If we observe $|0...0\ket$, we obtain no information about $s$ and we have to repeat the trial again.
Thus, in this case, the expected number of queries to obtain $s$ is equal to $[1+T_{\mbox{\scriptsize Q}}(1)]$.
Meanwhile if we observe $|s\ket$,
we obtain $s$ by a single query.
This consideration yields a relation
\beq
T_{\mbox{\scriptsize Q}}(1)=\frac{1}{2}[1+T_{\mbox{\scriptsize Q}}(1)]
+\frac{1}{2}\cdot 1,
\eeq
and we obtain $T_{\mbox{\scriptsize Q}}(1)=2$.

Let us evaluate $T_{\mbox{\scriptsize Q}}(2)$.
(We are given $\mbox{wt}(s)=m=2$.)
We can assume $s=110...0$ without losing generality.
If we observe the first register in the fifth step,
we obtain $000...0$, $010...0$, $100...0$, or $110...0$
as the binary string $k$ with probability $1/4$ respectively,
as shown in Eq.~(\ref{Q-algorithm-4th-step}).
We pay attention only to the first two bits of $k$, $k_{1}$ and $k_{2}$,
because the other bits (that is, $k_{3},...,k_{n}$) always hold zeros
as entries.
Each of $k_{1}$ and $k_{2}$ takes either zero or one with probability $1/2$
independently.
If $k_{i}=0$ for $i=1,2$,
we cannot determine an entry of $s_{i}$ and carry out another trial.
If $k_{i}=1$ for $i=1,2$,
we obtain $s_{i}=1$.

We have shown two methods for deriving $T_{\mbox{\scriptsize Q}}(1)$ before.
Here, we use the latter to evaluate $T_{\mbox{\scriptsize Q}}(2)$.
In the first trial, we obtain $(0,0)$, $(0,1)$, $(1,0)$, or $(1,1)$
for $(k_{1},k_{2})$ with probability $1/4$ respectively.
If we observe $(0,0)$, we obtain no information about $s$ and have to have another trial.
Thus, the expected number of queries for determining $s$ is equal to $[1+T_{\mbox{\scriptsize Q}}(2)]$.
If we observe $(0,1)$ or $(1,0)$, we can determine one of two bits
that have ones as entries in the string $s$.
Thus, the expected number of queries for obtaining $s$ is equal to $[1+T_{\mbox{\scriptsize Q}}(1)]$.
If we observe $(1,1)$, we obtain $s$ by a single query.
From this consideration, we can describe $T_{\mbox{\scriptsize Q}}(2)$ as
\beq
T_{\mbox{\scriptsize Q}}(2)=\frac{1}{4}[1+T_{\mbox{\scriptsize Q}}(2)]
+\frac{1}{4}[1+T_{\mbox{\scriptsize Q}}(1)]\cdot 2
+\frac{1}{4}\cdot 1.
\eeq
Using $T_{\mbox{\scriptsize Q}}(1)=2$, we obtain $T_{\mbox{\scriptsize Q}}(2)=8/3$.

For general $m(\geq 1)$, we have the following recurrence formula,
\beqa
T_{\mbox{\scriptsize Q}}(m)
&=&\frac{1}{2^{m}}
[(1+T_{\mbox{\scriptsize Q}}(m))
+{m\choose 1}(1+T_{\mbox{\scriptsize Q}}(m-1)) \non \\
&&\quad
+...+{m\choose m-1}(1+T_{\mbox{\scriptsize Q}}(1))+1] \non \\
&=&\frac{1}{2^{m}}
\sum_{h=0}^{m}{m\choose h}[1+T_{\mbox{\scriptsize Q}}(h)],
\lab{query-quantum-algorithm-m}
\eeqa
where $T_{\mbox{\scriptsize Q}}(0)=0$.
Using this formula, we can derive $T_{\mbox{\scriptsize Q}}(1)=2$, $T_{\mbox{\scriptsize Q}}(2)=8/3$, $T_{\mbox{\scriptsize Q}}(3)=22/7$,
and so on, in order from $m=1$.
From this discussion, we notice that $T_{\mbox{\scriptsize Q}}(m)$ does not depend on $n$.

It is difficult to derive a closed-form solution of $T_{\mbox{\scriptsize Q}}(m)$ from Eq.~(\ref{query-quantum-algorithm-m}).
Thus, we estimate $T_{\mbox{\scriptsize Q}}(m)$ roughly as follows.
Let us suppose that $\mbox{wt}(s)=m$ is very large.
We can assume $s=1...10...0$ for simplicity without losing generality.
(Hence, we assume $s_{i}=1$ for $i=1,...,m$ and $s_{i}=0$ for $i=m+1,...,n$.
Furthermore we assume $1\ll m<n$.)
In the fifth step of our algorithm,
we observe one of binary strings
$\{k:k_{i}\in\{0,1\}\mbox{ for $i=1,...,m$},\mbox{ and }k_{i}=0\mbox{ for $i=m+1,...,n$}\}$
at random.
Each of the first $m$ bits takes either zero or one as an entry with probability $1/2$ independently.
If we observe $k_{i}=1$ for $i=1,...,m$, we obtain $s_{i}=1$.
If we observe $k_{i}=0$ for $i=1,...,m$, we cannot determine $s_{i}$ and have to have another trial.

Here, let us suppose $m=2^{t}$.
In the observation of the first trial, half of the first $2^{t}$ bits (that is, about $2^{t-1}$ bits)
hold ones as entries,
and we put them on entries of $s$.
In the observation of the second trial, half of the rest undecided (that is, about $2^{t-2}$ bits)
hold ones as entries, and we put them on entries of $s$.
If we repeat this process $t$ times,
about one bit of $s$ is left undecided.
The expected number of queries for deciding a single bit is given by $T_{\mbox{\scriptsize Q}}(1)=2$.
Hence, we obtain $T_{\mbox{\scriptsize Q}}(2^{t})\sim 2+t$.
This result suggests the following approximate equation:
\beq
T_{\mbox{\scriptsize Q}}(m)\sim 2+\log_{2}m.
\lab{query-quantum-algorithm-approximation}
\eeq

Figure~\ref{Q-query} represents $T_{\mbox{\scriptsize Q}}(m)$ obtained by Eq.~(\ref{query-quantum-algorithm-m})
and its approximate value obtained by Eq.~(\ref{query-quantum-algorithm-approximation})
for $1\leq m\leq 500$.
It shows that Eq.~(\ref{query-quantum-algorithm-approximation}) is a good approximation.
We can conclude that the expected number of queries
$T_{\mbox{\scriptsize Q}}(m)$ is of order $\log m$.

\begin{figure}
\begin{center}
\includegraphics[scale=0.9]{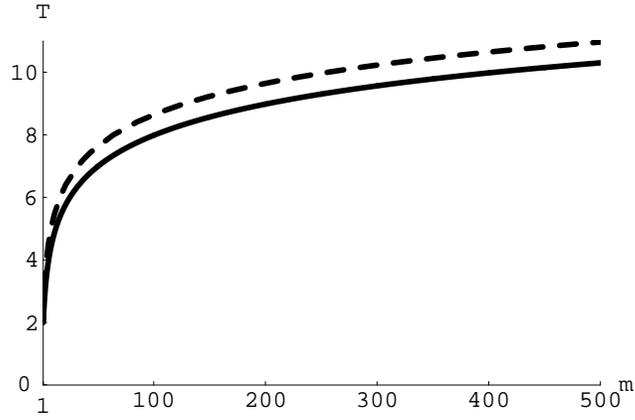}
\end{center}
\caption{The expected number of queries required by the quantum algorithm
defined in Sec.~\ref{problem-quantum-algorithm}
and its approximation.
A horizontal axis represents $m=\mbox{wt}(s)$, the Hamming weight of $s$,
and
a vertical axis represents $T$, the expected number of queries.
Both of $m$ and $T$ are dimensionless.
We set $1\leq m\leq 500$.
A solid curve shows $T_{\mbox{\scriptsize Q}}(m)$ obtained by Eq.~(\ref{query-quantum-algorithm-m}),
and a dashed curve shows an approximate value of $T_{\mbox{\scriptsize Q}}(m)$
obtained by Eq.~(\ref{query-quantum-algorithm-approximation}).}
\lab{Q-query}
\end{figure}

Our quantum algorithm has two features.
The first feature is as follows.
Our quantum algorithm gives us information about the binary string $s$
which determines the invariance of the function $f$,
although it does not tell us which value $f(x)$ takes for each input $x$.

In Sec.~\ref{problem-quantum-algorithm}, we show a concrete example of $f$
that takes $n=3$, $s=110$, and $\mbox{wt}(s)=m=2$.
Every input $x\in\{0,1\}^{3}$ is classified into one of four classes
according to its output $f(x)$,
as shown in Eq.~(\ref{classification-inputs}).
This classification is decided by $s$.
We can call it the global property of $f$.
By contrast, what value each $f(x)$ takes
(that is, which element of $\{0,1\}^{2}$
$f(000)$, $f(010)$, $f(100)$, and $f(110)$ take in Eq.~(\ref{classification-inputs})
respectively)
can be called the local property of $f$.
The function $f$ consists of the global property and the local properties.

Our quantum algorithm extracts only the global information of $f$.
We can find this feature in other quantum algorithms as well.

The second feature is as follows.
If we draw a network of quantum gates for our algorithm,
it is the same as that for Simon's algorithm
(see Fig.~\ref{qnetwork}).
Both algorithms differ only in promises of their oracles.
Our algorithm examines a function that has the invariance $f(x\wedge s)=f(x)$,
while Simon's algorithm examines a function that has the invariance $f(x\oplus s)=f(x)$.

\begin{figure}
\begin{center}
\includegraphics[scale=0.9]{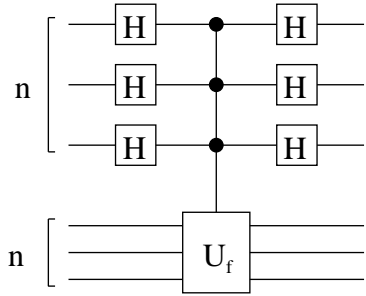}
\end{center}
\caption{A network of quantum gates for the quantum algorithm defined in Sec.~\ref{problem-quantum-algorithm}.
Simon's algorithm also works on this network.}
\lab{qnetwork}
\end{figure}

In Fig.~\ref{qnetwork}, our quantum algorithm seems to require
$(2n+1)$ quantum gates.
However, we can eliminate $2n$ Hadamard transformations $H$
by changing the initial state and the orthogonal basis for measurement.
Thus, a quantum gate that our algorithm essentially needs is only the oracle $U_{f}$.

\section{The classical lower bounds of the number of queries
for $\mbox{wt}(s)=1$ and $\mbox{wt}(s)=n-1$}
\lab{queries-classical-algorithm-lower-limit}
To show that the quantum algorithm introduced in Sec.~\ref{problem-quantum-algorithm}
is more efficient than any classical algorithm,
we need to know the lower bound of the number of queries required by an arbitrary classical algorithm.
However, in general, it is difficult to evaluate the classical lower bound.
In this section, we evaluate the classical lower bounds for $\mbox{wt}(s)=1$ and $\mbox{wt}(s)=n-1$ exactly.

\subsection{The case of $\mbox{wt}(s)=1$}
\lab{binary-search}
We discuss the case of $\mbox{wt}(s)=m=1$.
Let us think about the following classical algorithm.
For simplicity, we assume $n=2^{t}$.
We define $t$ strings of length $n$ as follows:
\beqa
a^{(1)}&=&0101...01, \non \\
a^{(2)}&=&00110011...0011, \non \\
\vdots && \non \\
a^{(t)}&=&0...01...1.
\eeqa
$a^{(l)}$ is a string of $2^{l-1}$ zeros alternating with a string of $2^{l-1}$ ones
for $l=1,...,t$.

We can decide the binary string $s=(s_{i})$ in the following way.
First, we compute $f(\mbox{\boldmath $0$})$ and $f(a^{(t)})$.
If $f(\mbox{\boldmath $0$})=f(a^{(t)})$,
there exists an only nonzero entry in the left half of the $n$-bit string $s$,
that is, $s_{i}=\delta_{ij}$ where $1\leq j\leq n/2$.
Meanwhile, if $f(\mbox{\boldmath $0$})\neq f(a^{(t)})$,
there exists the only nonzero entry
in the right half of the $n$-bit string $s$,
that is, $s_{i}=\delta_{ij}$ where $(n/2)+1\leq j\leq n$.

Here, for simplicity, we assume $f(\mbox{\boldmath $0$})=f(a^{(t)})$.
Next, we compute $f(a^{(t-1)})$.
If $f(\mbox{\boldmath $0$})=f(a^{(t-1)})$,
there exists the only nonzero entry in the first quarter of the $n$-bit string $s$
from the left side,
that is, $s_{i}=\delta_{ij}$ where $1\leq j\leq n/4$.
Meanwhile, if $f(\mbox{\boldmath $0$})\neq f(a^{(t-1)})$,
there exists the only nonzero entry in the second quarter of the $n$-bit string $s$
from the left side,
that is, $s_{i}=\delta_{ij}$ where $(n/4)+1\leq j\leq n/2$.

By repeating the above process,
we locate the only nonzero entry in the string $s$.
If we use the binary search explained above,
we can obtain $s$ by $(t+1)$ queries.
For example, when $s=10...0$, we obtain $s$ by answers of queries, $f(\mbox{\boldmath $0$})=f(a^{(t)})=...=f(a^{(1)})$.
When $s=010...0$, we obtain $s$ by answers of queries, $f(\mbox{\boldmath $0$})=f(a^{(t)})=...=f(a^{(2)})\neq f(a^{(1)})$.

For general $n$, the number of queries to obtain $s$ is given by
\beq
1+\lceil \log_{2}n \rceil,
\lab{binary-search-n=1}
\eeq
where $\lceil x \rceil$ denotes a unique integer $j$ such that $j-1<x\leq j$
for any real number $x$.

We show that this binary search is the most efficient algorithm of all classical algorithms.
We consider the information-theoretic lower bound of the number of classical queries.
(A discussion given here is concerned with an application of the coin-weighing problem \cite{coin-weighing,Terhal-Smolin}.
The coin-weighing problem is as follows:
``Suppose that we are given $n$ coins, one of which may be a forgery.
The forged coin is either too light or too heavy.
We are also given a balance on which we can place any of the coins we wish.
We want to determine whether the forgery exists or not, and if it exists,
we want to figure out which coin is false.
Ascertain the minimum number of uses of the balance to accomplish this task.'')

Let us count the number of possible functions for $f$.
We note that $f$ is a $2^{n-1}$-to-one function
because of the invariance $f(x\wedge s)=f(x)$,
where $\mbox{wt}(s)=m=1$.
(We explained this fact in Sec.~\ref{problem-quantum-algorithm}.)
Hence, we can rewrite $f$ as $f:\{0,1\}^{n}\rightarrow\{0,1\}$.
Then, $f$ is a surjection.
There are $n$ possible binary strings for the $n$-bit string $s$ because of $\mbox{wt}(s)=1$.
Moreover, we can divide the domain of $f$ (that is, $\{0,1\}^{n}$) into two subsets
as follows:
\beqa
X_{0}&=&\{a:a=y\wedge\overline{s},y\in\{0,1\}^{n}\}, \non \\
X_{1}&=&\{a\oplus s:a=y\wedge\overline{s},y\in\{0,1\}^{n}\}.
\eeqa
We have $|X_{0}|=|X_{1}|=2^{n-1}$, where $|X|$ denotes the number of elements in a set $X$.
Clearly, $\mbox{\boldmath $0$}=0...0\in X_{0}$ and $s\in X_{1}$.
We have the following relation:
\beq
f(x)=
\left\{
\begin{array}{ll}
f(\mbox{\boldmath $0$}) & \mbox{for $x\in X_{0}$} \\
f(s) & \mbox{for $x\in X_{1}$}
\end{array}
\right..
\eeq
When we think about the range of $f$ (that is, $\{0,1\}$),
we have two cases:
(1) $f(\mbox{\boldmath $0$})=0$ and $f(s)=1$;
(2) $f(\mbox{\boldmath $0$})=1$ and $f(s)=0$.
From the above discussion, we can conclude that there are $2n$ possible functions for $f$.

Let us suppose that these $2n$ functions are realized with equal probability,
\beq
P_{\alpha}=\frac{1}{2n} \mbox{ for } \alpha=1,...,2n,
\eeq
where $\alpha$ is an index of the functions.
Then, writing the amount of information that the problem holds as $S$,
which can be called entropy, it is given by
\beq
S=-\sum_{\alpha}P_{\alpha}\log_{2}P_{\alpha}=\log_{2}(2n)
=1+\log_{2}n.
\lab{entropy-f}
\eeq
By contrast, writing the amount of information retrieved by a single query as $A$,
it is given by
\beq
A=-2\cdot\frac{1}{2}\cdot\log_{2}\frac{1}{2}=1.
\eeq
This is because the query has two possible answers,
zero and one, as values of $f(x)$ (for $x\in\{0,1\}^{n}$),
and both the answers appear with probability $1/2$ respectively.
Hence, the lower bound of the number of classical queries for solving the problem is given by
\beq
\lceil (S/A) \rceil=1+\lceil\log_{2}n\rceil.
\lab{minimum-query-number}
\eeq

The number of queries given in Eq.~(\ref{minimum-query-number}) is equal to
the number of queries for the binary search given in Eq.~(\ref{binary-search-n=1}).
Thus, we can conclude that the binary search explained before is the most efficient algorithm of all classical algorithms.
Contrastingly, the expected number of queries required by the quantum algorithm defined
in Sec.~\ref{problem-quantum-algorithm}
is given by $T_{\mbox{\scriptsize Q}}(1)=2$.
Hence, we can conclude that our quantum algorithm is more efficient than any classical algorithm.

Here, we note the following fact.
Equation~(\ref{entropy-f}) shows that the amount of information that the problem holds is equal to $(1+\log_{2}n)$.
Meanwhile, there are $n$ possible binary strings for $s$ that represents the global information of $f$.
There also exist two cases, $(f(\mbox{\boldmath $0$}),f(s))=(0,1)$ and $(1,0)$,
which represent the local information of $f$.
Hence, the global information of the function $f$ amounts to $\log_{2}n$
and the local information of $f$ amounts to $\log_{2}2=1$.

If we want to know only $s$,
we can expect that the minimum number of queries will be given by $\lceil\log_{2}n\rceil$.
However, the binary search cannot distinguish the global information and the local information.
(The binary search cannot extract only the global information of the oracle.)
Therefore, the lower bound of the number of classical queries is equal to $(1+\lceil\log_{2}n\rceil)$.

\subsection{The case of $\mbox{wt}(s)=n-1$}
\lab{sequential-search}
We consider the case of $\mbox{wt}(s)=m=n-1$.
Then, the promise $f(x\wedge s)=f(x)$ is rewritten as
\beq
f(x)=f(x\oplus \overline{s})\quad\mbox{where $\mbox{wt}(\overline{s})=1$}.
\lab{invariance-condition-oplus}
\eeq
$f$ is two-to-one.
These facts can be seen in the example that holds $n=3$, $s=110$, and $\mbox{wt}(s)=2$
in Eq.~(\ref{classification-inputs}).

In the case of Eq.~(\ref{invariance-condition-oplus}),
the best classical algorithm is as follows.
There are $n$ possible binary strings for $\overline{s}$.
(The Hamming weight of each possible string is equal to one.)
We examine whether or not these $n$ binary strings satisfy Eq.~(\ref{invariance-condition-oplus})
one by one in order.
This is a sequential search.

We define the following $n$ strings of length $n$:
\beqa
b^{(1)}&=&10...0, \non \\
b^{(2)}&=&010...0, \non \\
\vdots && \non \\
b^{(n)}&=&0...01,
\lab{sequential-search-bit-strings}
\eeqa
where $b^{(i)}=(b^{(i)}_{j})=(\delta_{ij})$ for $i,j=1,...,n$.
$b^{(i)}$ is a string whose Hamming weight is equal to one.

First, we compute $f(\mbox{\boldmath $0$})$.
Next, we compute $f(b^{(1)})$.
If $f(b^{(1)})=f(\mbox{\boldmath $0$})$,
we obtain $s=\overline{b^{(1)}}=01...1$.
If $f(b^{(1)})\neq f(\mbox{\boldmath $0$})$,
we compute $f(b^{(2)})$.
In this way, we compute $f(b^{(i)})$ for $i=1,2,3,...$ in order,
and we obtain $s=\overline{b^{(i)}}$ when $f(b^{(i)})=f(\mbox{\boldmath $0$})$.

We evaluate the expected number of queries
to obtain $s$.
We assume that $n$ possible strings for $s$ appear with equal probability.
(The Hamming weight of each possible $n$-bit string is given by $(n-1)$.)
If $s=01...1$, the number of queries is equal to two.
If $s=101...1$, the number of queries is equal to three.
In contrast, if $s=1...10$, the number of queries is equal to $n$
because $s$ is decided by $f(b^{(n-1)})\neq f(\mbox{\boldmath $0$})$.
Thus, the expected number of queries is given by
\beq
\frac{1}{n}(2+3+...+n+n)
=-\frac{1}{n}+\frac{3}{2}+\frac{n}{2}.
\lab{sequential-search-m=n-1}
\eeq

Contrastingly, from Eq.~(\ref{query-quantum-algorithm-approximation}),
the expected number of queries required by the quantum algorithm defined
in Sec.~\ref{problem-quantum-algorithm}
is given by
\beq
T_{\mbox{\scriptsize Q}}(n-1)\sim 2+\log_{2}(n-1).
\eeq
Hence, our quantum algorithm is more efficient than any classical algorithm in the case of $\mbox{wt}(s)=n-1$.

\section{The number of queries of classical algorithms for $2\leq\mbox{wt}(s)\leq n-2$}
\lab{queries-classical-algorithm-between-2-(n-2)}
In Sec.~\ref{queries-classical-algorithm-lower-limit},
we evaluate the lower bounds of the number of classical queries for
$\mbox{wt}(s)=1$ and $\mbox{wt}(s)=n-1$,
and we show that the quantum algorithm introduced in Sec.~\ref{problem-quantum-algorithm}
is more efficient than any classical algorithm in those cases.
However, it is difficult to evaluate the classical lower bound for $2\leq\mbox{wt}(s)\leq n-2$.
Thus, in this section, we introduce two typical classical algorithms
and evaluate the number of queries for each of them.
We compare the efficiency of our quantum algorithm with that of the two classical algorithms.

\subsection{An application of the binary search}
Let us consider the first typical classical algorithm as follows.
In Sec.~\ref{binary-search},
we show that the binary search is the most efficient algorithm
of all the classical algorithms for $\mbox{wt}(s)=1$.
We adapt this method to the case of $2\leq\mbox{wt}(s)\leq n-2$.
We locate nonzero entries in the string $s$
by repeating the binary search $m$ times,
where $m=\mbox{wt}(s)$.

We use the following fact.
We suppose that we do not know $s$ except that $s$ is an $n$-bit string and $\mbox{wt}(s)=m$.
We define an $n$-bit string $u=(u_{i})$ as
\beq
u_{i}=
\left\{
\begin{array}{ll}
0 & \mbox{for $1\leq i\leq l(<n)$} \\
1 & \mbox{for $l+1\leq i\leq n$}
\end{array}
\right..
\eeq
Comparing $f(\mbox{\boldmath $0$})$ and $f(u)$,
we obtain one of two cases:
(1) if $f(\mbox{\boldmath $0$})=f(u)$, $s_{l+1}=s_{l+2}=...=s_{n}=0$;
(2) if $f(\mbox{\boldmath $0$})\neq f(u)$, at least one of $s_{l+1}$, $s_{l+2}$, ..., $s_{n}$
holds a nonzero entry ($s_{i}=1$ for some $i\in\{l+1,l+2,...,n\}$).

In the concrete, we decide $s$ as follows.
We assume $n=2^{t}$ for simplicity.
First, we compute $f(\mbox{\boldmath $0$})$.
Next, we compute $f(u^{(1)})$, where $u^{(1)}=(u^{(1)}_{i})$ is given by
\beq
u^{(1)}_{i}=
\left\{
\begin{array}{ll}
0 & \mbox{for $i=1,2,...,n/2$} \\
1 & \mbox{for $i=(n/2)+1,(n/2)+2,...,n$}
\end{array}
\right..
\lab{application-binary-search-1}
\eeq
If $f(\mbox{\boldmath $0$})=f(u^{(1)})$,
we obtain $s_{(n/2)+1}=s_{(n/2)+2}=...=s_{n}=0$.
This implies all of ones in entries exist in the left half of the string $s$.
By contrast, if $f(\mbox{\boldmath $0$})\neq f(u^{(1)})$,
at least one of $s_{(n/2)+1}$, $s_{(n/2)+2}$, .., $s_{n}$
holds a nonzero entry.
This implies at least one bit holds a nonzero entry in the right half of the string $s$.

Here, let us suppose $f(\mbox{\boldmath $0$})=f(u^{(1)})$.
We define $u^{(2)}=(u^{(2)}_{i})$, where
\beq
u^{(2)}_{i}=
\left\{
\begin{array}{ll}
0 & \mbox{for $i=1,2,...,n/4$ and $i=(n/2)+1,(n/2)+2,...,n$} \\
1 & \mbox{for $i=(n/4)+1,(n/4)+2,...,n/2$}
\end{array}
\right..
\eeq
We compare $f(\mbox{\boldmath $0$})$ and $f(u^{(2)})$.
From this act, we find which quarter of the string $s$ has at least a nonzero entry of a bit,
the first quarter or the second quarter from the left side.

Next, let us suppose $f(\mbox{\boldmath $0$})\neq f(u^{(1)})$.
We define $u^{(2)}=(u^{(2)}_{i})$, where
\beq
u^{(2)}_{i}=
\left\{
\begin{array}{ll}
0 & \mbox{for $i=1,2,...,3(n/4)$} \\
1 & \mbox{for $i=3(n/4)+1,3(n/4)+2,...,n$}
\end{array}
\right..
\eeq
We compare $f(\mbox{\boldmath $0$})$ and $f(u^{(2)})$.
From this act, we find which quarter of the string $s$ has at least a nonzero entry of a bit,
the third quarter or the forth quarter from the left side.

By the repetition of this process,
we can locate a nonzero entry of a bit in the string $s$ by $1+t=1+\log_{2}n$ queries.
(This is the binary search.)
The string $s$ includes $m$ nonzero entries because of $\mbox{wt}(s)=m$.
We suppose that we locate one of these nonzero entries by the above method.
Then, the problem is simplified.
A new problem is to locate $(m-1)$ nonzero entries in an $(n-1)$-bit string.

For example, let us suppose that the right end of the string $s$ is given by $s_{n}=1$
and we have located it first.
We can locate another nonzero entry of $s$ as follows.
We define $u'^{(1)}=(u'^{(1)}_{i})$ where
\beq
u'^{(1)}_{i}=
\left\{
\begin{array}{ll}
0 & \mbox{for $i=1,2,...,n/2$ and $i=n$} \\
1 & \mbox{for $i=(n/2)+1,(n/2)+2,...,n-1$}
\end{array}
\right.,
\eeq
as a substitute of $u^{(1)}$ in Eq.~(\ref{application-binary-search-1}).
Thus, $u'^{(1)}$ has a form $0...01...10$.
Because $s_{n}=1$ is detected,
we put a zero in the $n$th bit of $u'^{(1)}$,
put zeros in the first half of $(n-1)$ undecided bits,
and put ones in the second half of them.

If $f(\mbox{\boldmath $0$})=f(u'^{(1)})$,
we obtain $s_{(n/2)+1}=s_{(n/2)+2}=...=s_{n-1}=0$.
This implies that there are $(m-1)$ ones in the left half of entries of $s$.
($(m-1)$ bits of $s_{1}$, $s_{2}$, ..., $s_{n/2}$ have ones as entries.)
By contrast, if $f(\mbox{\boldmath $0$})\neq f(u'^{(1)})$,
at least one of $s_{(n/2)+1}$, $s_{(n/2)+2}$, ..., $s_{n-1}$
has a nonzero entry.

As shown above,
if we apply the binary search to unknown $(n-1)$ bits of $s$,
we can find the second nonzero entry of $s$ by $\lceil \log_{2}(n-1)\rceil$ queries.
Hence, if we write the number of queries to obtain $s$ by this classical algorithm as $T^{n}_{\mbox{\scriptsize CB}}(m)$,
it is given by
\beq
T^{n}_{\mbox{\scriptsize CB}}(m)=1+\sum_{h=1}^{m}\lceil \log_{2}(n-h+1)\rceil
\quad
\mbox{for $m=1,...,n-1$}.
\lab{binary-search-m}
\eeq
The subscripts C and B of $T^{n}_{\mbox{\scriptsize CB}}(m)$ stand for ``classical'' and ``binary'', respectively.
Moreover, we note that Eq.~(\ref{binary-search-m}) gives us
$T^{n}_{\mbox{\scriptsize CB}}(1)=1+\lceil \log_{2}n\rceil$
and it corresponds with Eq.~(\ref{binary-search-n=1}).

\subsection{An application of the sequential search}
Let us consider the second typical classical algorithm as follows.
In Sec.~\ref{sequential-search},
we show that the sequential search is the most efficient algorithm
of all the classical algorithms for $\mbox{wt}(s)=n-1$.
We adapt this method to the case of $2\leq\mbox{wt}(s)\leq n-2$.
We use the $n$ strings of length $n$, $b^{(i)}$ (for $i=1,...,n$),
defined in Eq.~(\ref{sequential-search-bit-strings}) again.

First, we compute $f(\mbox{\boldmath $0$})$.
Next, we compute $f(b^{(1)})$.
If $f(b^{(1)})=f(\mbox{\boldmath $0$})$,
we obtain $s_{1}=0$, where $s_{1}$ is the first bit of the string $s$.
By contrast, if $f(b^{(1)})\neq f(\mbox{\boldmath $0$})$,
we obtain $s_{1}=1$.
Likewise, computing $f(b^{(i)})$ and applying the following rule
\beq
\left\{
\begin{array}{lll}
f(b^{(i)})=f(\mbox{\boldmath $0$}) & \rightarrow & s_{i}=0 \\
f(b^{(i)})\neq f(\mbox{\boldmath $0$}) & \rightarrow & s_{i}=1
\end{array}
\right.
\eeq
to it for $i=1,2,...,n$ in order,
we decide entries of the string $s$ one by one from the first bit.
We are given $\mbox{wt}(s)=m$ beforehand.
Thus, when $m$ nonzero bits appear in the string $s$ in the middle of the above process,
we can decide the whole $s$ immediately and finish the task.

Let us write the expected number of queries required by this algorithm for $\mbox{wt}(s)=m$
as $T^{n}_{\mbox{\scriptsize CS}}(m)$.
The subscripts C and S of $T^{n}_{\mbox{\scriptsize CS}}(m)$ stand for ``classical'' and ``sequential'', respectively.
We can find the following properties of $T^{n}_{\mbox{\scriptsize CS}}(m)$ instantly.
Clearly, $T^{n}_{\mbox{\scriptsize CS}}(m)\leq n$.
(Because we are given $\mbox{wt}(s)=m$, we can always decide the whole $s$ certainly by
computing $f(\mbox{\boldmath $0$})$, $f(b^{(1)})$, ..., and $f(b^{(n-1)})$.)
Moreover, we have $T^{n}_{\mbox{\scriptsize CS}}(m)=T^{n}_{\mbox{\scriptsize CS}}(n-m)$.
This is because specifying the string $s$ that has $\mbox{wt}(s)=m$
corresponds to not only locating $m$ nonzero bits but also locating $(n-m)$ entries
that have zeros.
Furthermore, we have obtained
\beq
T^{n}_{\mbox{\scriptsize CS}}(n-1)=T^{n}_{\mbox{\scriptsize CS}}(1)
=-\frac{1}{n}+\frac{3}{2}+\frac{n}{2}
\lab{sequential-search-m=1}
\eeq
in Eq.~(\ref{sequential-search-m=n-1}) already.

As a concrete example, we calculate $T^{n}_{\mbox{\scriptsize CS}}(2)$.
When $\mbox{wt}(s)=2$, there are $n\choose 2$ possible strings for $s$.
The string $s=110...0$ requires the fewest queries among them.
It requires three queries.
Strings that require four queries are $s=1010...0$ and $s=0110...0$.
Likewise, if $3\leq l\leq n-2$, there exist $(l-2)$ possible strings
that are specified just with $l$ queries.

However, we cannot apply the similar discussion to strings that is specified with $(n-1)$ queries.
First, $(n-3)$ strings, $10...0100$, $010...0100$, ..., $0...01100$,
whose last three bits are given by $100$,
are specified by $(n-1)$ queries.
Moreover, a string $0...011$ can be specified by $(n-1)$ queries as well.
(The string $0...011$ is specified when the first $(n-2)$ zeros are located
as entries.)
Hence, there exist $[(n-3)+1]$ strings that can be specified just by $(n-1)$ queries.

Strings specified by $n$ queries are as follows.
First, $(n-2)$ strings, $10...010$, $010...010$, ..., $0...0110$,
whose last two bits are given by $10$,
are specified by $n$ queries.
Furthermore, $(n-2)$ strings, $10...01$, $010...01$, ..., $0...0101$,
whose last two bits are given by $01$,
are specified by $n$ queries as well.
Thus, the number of strings that are specified just by $n$ queries is $[(n-2)+(n-2)]$.

Hence, we obtain
\beq
T^{n}_{\mbox{\scriptsize CS}}(2)=\frac{1}{{n\choose 2}}[\sum_{h=3}^{n}h(h-2)+(n-1)+n(n-2)].
\lab{sequential-search-m=2}
\eeq
From similar discussion, for $1\leq m\leq n-1$,
we obtain
\beq
T^{n}_{\mbox{\scriptsize CS}}(m)=\frac{1}{{n\choose m}}[\sum_{h=m+1}^{n}h{h-2\choose m-1}+\sum_{h=0}^{m-1}(n-h){n-2-h\choose m-1-h}].
\lab{sequential-search-m}
\eeq
The definition of $T^{n}_{\mbox{\scriptsize CS}}(m)$ in Eq.~(\ref{sequential-search-m})
includes the expression of $T^{n}_{\mbox{\scriptsize CS}}(n-1)$ given in Eq.~(\ref{sequential-search-m=1})
and that of $T^{n}_{\mbox{\scriptsize CS}}(2)$ given in Eq.~(\ref{sequential-search-m=2}).
Moreover, Eq.~(\ref{sequential-search-m}) satisfies the relation $T^{n}_{\mbox{\scriptsize CS}}(m)=T^{n}_{\mbox{\scriptsize CS}}(n-m)$.

Figure~\ref{QandC-query} shows $T_{\mbox{\scriptsize Q}}(m)$ defined in Eq.~(\ref{query-quantum-algorithm-m}),
$T^{n}_{\mbox{\scriptsize CB}}(m)$ defined in Eq.~(\ref{binary-search-m}),
and $T^{n}_{\mbox{\scriptsize CS}}(m)$ defined in Eq.~(\ref{sequential-search-m})
for $n=200$ and $1\leq m\leq 199$.
A horizontal axis represents $m$, the Hamming weight of $s$, and a vertical axis represents $T$,
the expected number of queries.
$T$ on the vertical axis is represented on a logarithmic scale.
Figure~\ref{QandC-query} shows that the quantum algorithm defined in Sec.~\ref{problem-quantum-algorithm}
is more efficient than two classical algorithms
(that is, the applications of the binary search and the sequential search)
discussed in this section.

\begin{figure}
\begin{center}
\includegraphics[scale=0.9]{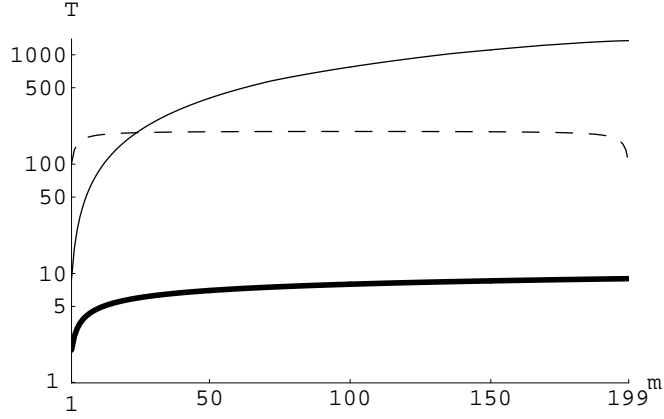}
\end{center}
\caption{The expected number of queries for solving the problem
defined in Sec.~\ref{problem-quantum-algorithm}
with $n=200$ and $1\leq m\leq 199$.
A horizontal axis represents $m=\mbox{wt}(s)$, the Hamming weight of $s$, 
and a vertical axis represents $T$, the expected number of queries.
Both $m$ and $T$ are dimensionless.
$T$ on the vertical axis is represented on a logarithmic scale.
A thick solid curve represents $T_{\mbox{\scriptsize Q}}(m)$ defined in Eq.~(\ref{query-quantum-algorithm-m})
(the quantum algorithm introduced in Sec.~\ref{problem-quantum-algorithm}),
a thin solid curve represents $T^{n}_{\mbox{\scriptsize CB}}(m)$ defined in Eq.~(\ref{binary-search-m})
(the application of the binary search),
and a thin dashed curve represents $T^{n}_{\mbox{\scriptsize CS}}(m)$ defined in Eq.~(\ref{sequential-search-m})
(the application of the sequential search).}
\lab{QandC-query}
\end{figure}

\section{A quantum algorithm for examining an OR-mask invariant oracle}
\lab{OR-quantum-algorithm}
In the previous sections,
we consider the problem that is to find the binary string $s$ by querying an oracle $f$
that has the invariance
$f(x\wedge s)=f(x)$.
In this section, we consider a similar problem
that is to find $s$ by querying an oracle $g$ that has invariance 
$g(x\vee s)=g(x)$.

The problem is given explicitly as follows:
``Suppose that we are given a function
$g:\{0,1\}^{n}\rightarrow\{0,1\}^{n}$.
We are promised that there exists an $n$-bit string $s$ such that
$\forall x,y\in\{0,1\}^{n}$
($g(x)=g(y)$ if and only if
$x\vee s=y\vee s$).
We do not know $s$ except that we are given $\mbox{wt}(s)=n-m$.
Find $s$.''

If $\mbox{wt}(s)=0$ or $n$, $s$ is trivial.
Thus we assume $s\neq \mbox{\boldmath $0$},\mbox{\boldmath $1$}$.
$g$ is a $2^{n-m}$-to-one function.

We can solve the above problem by a quantum algorithm that is similar to the algorithm
introduced in Sec.~\ref{problem-quantum-algorithm}.
Let us carry out the steps 1, 2, 3, and 4
in the algorithm discussed in Sec.~\ref{problem-quantum-algorithm}.
First, we note the following fact.
$\forall x,y\in\{0,1\}^{n}$,
$g(x)=g(y)$ if and only if
$x\vee s=y\vee s$,
and $\mbox{wt}(s)=n-m$ is given.
Thus, $g$ is a $2^{n-m}$-to-one function.
This is because the number of bits that hold ones in the string $s$
is equal to $(n-m)$
and the function $g$ does not depend on these $(n-m)$ bits.
Hence, we can classify $2^{n}$ inputs of $g$
(that is, $x\in\{0,1\}^{n}$)
into $2^{m}$ classes according to values that they take as outputs of $g$
(that is, $g(x)$).
The number of inputs in each class is equal to $2^{n-m}$.

From this consideration, we can rewrite Eq.~(\ref{quantum-algorithm-3rd-step-register})
that is obtained after the third step of the algorithm as follows:
\beq
\frac{1}{\sqrt{2^{n}}}\sum_{l:l=x\wedge\overline{s},x\in\{0,1\}^{n}}
(\sum_{a:a=y\wedge s,y\in\{0,1\}^{n}}|a\oplus l\ket)
|g(l)\ket.
\lab{OR-problem-3rd-step}
\eeq
In Eq.~(\ref{OR-problem-3rd-step}),
the binary string $l$ has zeros in entries where $\overline{s}$ has zeros,
and $l$ has either zeros or ones at random in entries where $\overline{s}$ has ones.
Thus, there are $2^{m}$ possible strings for $l$.
Meanwhile, the binary string $a$ has zeros in entries where $s$ has zeros,
and $a$ has either zeros or ones at random in entries where $s$ has ones.
Thus, there are $2^{n-m}$ possible strings for $a$.

Next, we apply $H$ to each qubit of the first register in Eq.~(\ref{OR-problem-3rd-step})
for the fourth step.
Here, we consider only the $n$ qubits of the first register,
$(1/\sqrt{2^{n-m}})\sum_{a}|a\oplus l\ket$.
We permute these $n$ qubits,
so that ones of the string $s$ move to the left side and zeros of $s$ move to the right side.
By this permutation,
the state of the first register is rewritten as Eq.~(\ref{Q-algorithm-3rd-step-1st-register}).
$H^{\otimes n-m}$ transforms the superposition of states $|a'\ket$ to $|0\ket^{n-m}$.
Thus, Eq.~(\ref{OR-problem-3rd-step}) is transformed to the following state:
\beq
\frac{1}{2^{m}}\sum_{l:l=x\wedge\overline{s},x\in\{0,1\}^{n}}
[\sum_{k:k=y\wedge\overline{s},y\in\{0,1\}^{n}}
(-1)^{l\cdot k}|k\ket]|g(l)\ket.
\eeq

Then, we observe the first register in the basis $\{|x\ket:x\in\{0,1\}^{n}\}$.
By this observation, we obtain a binary string $k$,
where $k=y\wedge \overline{s}$ and $y\in\{0,1\}^{n}$.
There are $2^{m}$ possible strings for $k$,
and $k$ takes one of them at random.
$\overline{k}$ has either zeros or ones at random in entries where $s$ has zeros,
and $\overline{k}$ has ones in entries where $s$ has ones.
Thus, if we repeat the trial with observing a string $\overline{k}$
and perform the bitwise AND to observed strings $\overline{k}$ again and again,
we will obtain $s$ eventually.
(Suppose we obtain $k$.
If $\mbox{wt}(\overline{k})=n-m$, we let $s=\overline{k}$.
If $\mbox{wt}(\overline{k})>n-m$, we rewrite $k$ as $k_{old}$,
carry out the trial again,
obtain a new observed results $k_{new}$,
and have $\overline{k}=\overline{k_{old}}\wedge\overline{k_{new}}$.
We repeat this procedure.)

Clearly, the expected number of queries to obtain $s$ by this quantum algorithm
is equal to $T_{\mbox{\scriptsize Q}}(m)$ defined in Eq.~(\ref{query-quantum-algorithm-m}).
Furthermore, when we think about the efficiency of classical algorithms for finding $s$ such that
$g(x\vee s)=g(x)$,
we can have discussion similar to that held in Secs.~\ref{queries-classical-algorithm-lower-limit}
and \ref{queries-classical-algorithm-between-2-(n-2)}.

\section{Discussion}
\lab{discussions}
In this paper, we discuss the problem that is to find $s$ with querying an oracle,
where the oracle represents a function that has the invariance,
$f(x\wedge s)=f(x)$.
The quantum algorithm proposed in this paper is more efficient
than any classical algorithm for $\mbox{wt}(s)=1$ and $\mbox{wt}(s)=n-1$.
($n$ denotes the number of bits in an input of $f$.)

Our quantum algorithm requires $O(1)$ queries on average for $\mbox{wt}(s)=1$,
while any classical algorithm needs at least of the order of $\log n$ queries.
Likewise, our algorithm requires $O(\log n)$ queries on average for $\mbox{wt}(s)=n-1$,
while any classical algorithm needs at least of the order of $n$ queries.
(In both cases, our quantum algorithm is faster than any classical algorithm.)
However, in general, researchers' motivation for studying the quantum computation
is to solve a certain problem in quantum polynomial time in $n$
rather than classical exponential time in $n$.
(We cannot find an exponential gap in $n$
between our quantum algorithm and the best classical algorithm.)
From this point of view,
our quantum algorithm seems not to have a remarkable complexity theoretic advantage.
However, our algorithm makes good use of properties of quantum mechanics,
that is, the principle of superposition and its interference,
and entanglement.
Thus, we can say that our algorithm is one of genuine quantum algorithms.
Moreover, as mentioned in Sec.~\ref{queries-quantum-algorithm},
our quantum algorithm neglects the local properties of the oracle
and extracts only the global property of the oracle efficiently.

B.M.~Terhal and J.A.~Smolin have proposed a quantum algorithm for solving the binary search problem
with a single query \cite{Terhal-Smolin}.
By contrast, the classical lower bound of queries for this problem is equal to $\log_{2}n$,
where $n$ denotes the number of bits in an input of an oracle.
From a viewpoint of the complexity,
our algorithm resembles 
B.M.~Terhal and J.A.~Smolin's algorithm.

Simon's algorithm finds $s$ in polynomial time by querying the oracle $f$ that has the invariance
$f(x)=f(x\oplus s)$.
By contrast, any classical computer takes exponential time to find $s$.
However, the running time of Simon's algorithm is evaluated in the expected sense.
Hence, there exists a remote but finite possibility that Simon's algorithm needs exponential time
for finding $s$.

G.~Brassard, P.~H{\o}yer, T.~Mihara, and S.C.~Sung
have discussed quantum algorithms that are guaranteed to solve Simon's problem in polynomial
time in the worst case \cite{Brassard-Hoyer-Mihara-Sung}.
As mentioned in Sec.~\ref{queries-quantum-algorithm},
our quantum algorithm is an application of Simon's algorithm, and we evaluate its running time
(that is, the number of queries) in the expected sense.
Thus, it may be interesting to study a quantum algorithm that solves the problem discussed in this paper
more efficiently in the worst case than any classical algorithm.

\bigskip
\noindent
{\bf \large Acknowledgement}
\smallskip

We thank M.~Okuda for valuable discussion and encouragement.

\end{document}